\documentclass[aps,prl,twocolumn,superscriptaddress,floatfix,showpacs]{revtex4}

\usepackage{float}
\usepackage{placeins}
\usepackage{amssymb,amsmath}
\usepackage{graphicx}
\usepackage[colorlinks=true,linkcolor=blue,citecolor=blue,urlcolor=blue]{hyperref}

\begin{document}

\title{Application of a small oscillating magnetic field to reveal the peak effect in the resistivity of Nb$_3$Sn}

\author{M.\ Reibelt}
\email[]{reibelt@physik.uzh.ch}
\author{A.\ Schilling}
\affiliation{Physik-Institut University of Zurich, Winterthurerstrasse 190, CH-8057 Zurich, Switzerland}
\author{N.\ Toyota}
\affiliation{Physics Department, Graduate School of Science, Tohoku University, 980-8571 Sendai, Japan}

\begin{abstract}
By the application of a small oscillating magnetic field parallel to the main magnetic field and perpendicular to the transport current, we were able to unveil the peak effect in the resistivity data of Nb$_3$Sn near the upper critical field $H_{c2}$. We investigated the dependence of this effect on the frequency and the amplitude of the oscillating magnetic field and show that the used technique can be more sensitive to detect the peak effect in a certain range of temperatures and magnetic fields than conventional magnetization measurements.
\end{abstract}

\maketitle

\section{INTRODUCTION}

\begin{figure}
\includegraphics[width=75mm,totalheight=200mm,keepaspectratio]{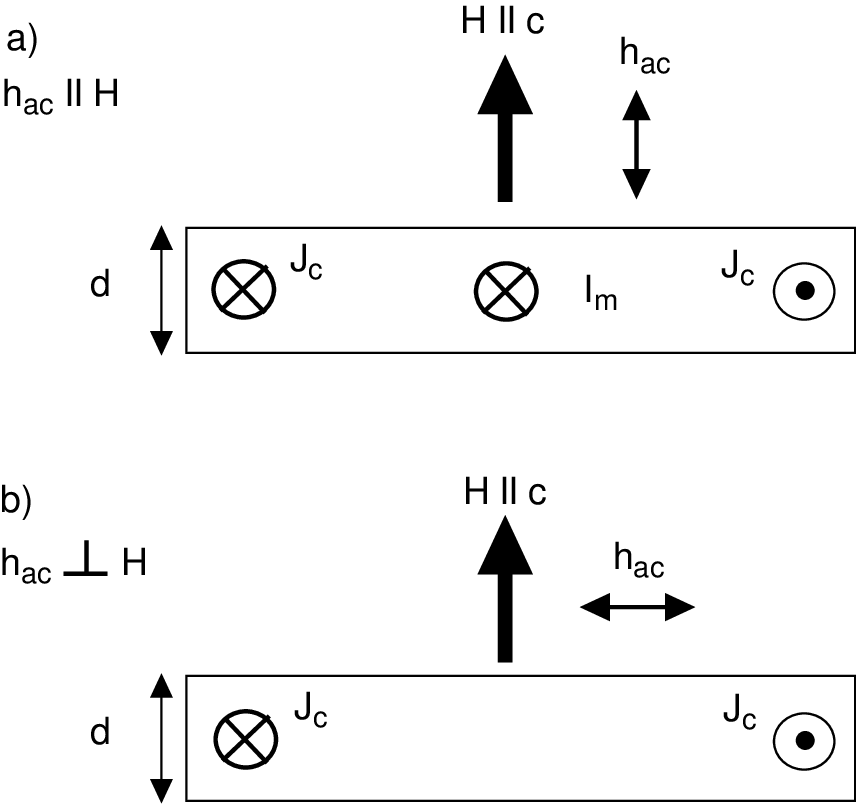}
\caption{Schematic drawing of the configurations \textbf{(a)} $h_{ac} \parallel H$ plus a transport current $I_{m}$ \ ; \ \textbf{(b)} $h_{ac} \perp H$, with no transport current . The sheet current $J_c$ generates the diamagnetic response of the superconductor. \label{fig.PRB0}}
\end{figure}

Vortex matter in type-II superconductors exhibits a huge variety of interesting effects, the investigation of which can provide information on the vortex dynamics and flux-line lattice (FLL) structures \cite{Blatter1994}. For example, the occurrence of a maximum in the critical current as the external magnetic field $H$ is varied near the upper-critical magnetic field $H_{c2}$ of certain hard superconductors, is known as the so-called \emph{peak effect}. At temperatures close to the critical temperature $T_c$ the flux-line lattice softens, and the magnetic-flux lines can adjust better to the pinning centers, which in turn leads to a higher pinning strength and thereby to an increase in the critical current. An increased pinning strength manifests itself as a "bubblelike" feature in dc magnetization-hysteresis data that results from an enhanced irreversibility and consequently a large difference between the two branches (increasing and decreasing magnetic field) of corresponding $M(H)$ data at the peak effect. The peak effect has been observed in different physical quantities, such as ac susceptibility \cite{Marchevsky2001February,Adesso2006March}, dc magnetization \cite{Lortz2007March}, and even in electrical-transport measurements \cite{Higgins1996,Chaudhary2001,Meier-Hirmer1985January,Wordenweber1986March,Henderson1996September,Daniilidis2007,Huebener1970February,Sato1997,Langan1997}.\\
\indent According to recent theories \cite{Brandt1986November,Brandt1994April,Brandt1996March,Brandt1996August,Mikitik2000September,Mikitik2001August,Brandt2002July,Brandt2004,Brandt2004a,Brandt2004b,Brandt2007}, transversal and also longitudinal "flux-line shaking" (relative to the major screening-sheet-current direction inside the superconducting sample) with an oscillating magnetic field ("shaking field") $h_{ac}$ can cause flux lines to "walk" through the superconductor, which leads to a reduction of nonequilibrium effects in the current distribution and thereby to an equilibration of the FFL. This technique has been successfully applied, for example, for resistivity measurements \cite{Cape1968February}, torque magnetometry \cite{Willemin1998September,Weyeneth2009}, local magnetization measurements \cite{Avraham2001May}, and also for specific-heat investigations \cite{Lortz2006September}.\\
\indent We have measured the resistivity of a Nb$_3$Sn single crystal while applying a shaking field $h_{ac}$ parallel to the main dc magnetic field $H$ (so-called $h_{ac} \parallel H$ configuration) and in the presence of a transport current oriented perpendicular to $H$. This configuration is different from, e.g., the setup of Cape and Silvera \cite{Cape1968February} who applied the small shaking field \emph{perpendicular} to the main dc magnetic field (so-called $h_{ac} \perp H$ configuration), a geometry that has been extensively theoretically studied by Brandt \emph{et al.}~\cite{Brandt1986November,Brandt1994April,Brandt1996March,Brandt1996August,Mikitik2000September,Brandt2002July,Brandt2004,Brandt2004a,Brandt2004b,Brandt2007}. In Fig.~\ref{fig.PRB0} we show a schematic view of the two configurations $h_{ac} \parallel H$ and $h_{ac} \perp H$, respectively. Only a very few experiments have been reported in the literature with a transport current in combination with a shaking field \cite{Huebener1971,Huebener1971a,Huebener1972,Risse1997June}. A configuration that is comparable to ours and with a similar shaking technique was used by Huebener and Rowe \cite{Huebener1971} to study lead films and by Risse \emph{et al.}~\cite{Risse1997June} who investigated amorphous Mo$_3$Si films with a similar shaking technique. \\
\indent The magnetic response of a flat regularly shaped superconducting sample in
the mixed state to an oscillatory magnetic field
component parallel to the sample surface is strictly symmetric, i.e., magnetic-flux lines enter and leave the sample on opposite edges in the same way. This geometrical symmetry is lifted as soon as an additional transport current is applied. As a result, the formation of a dc electric field parallel to the transport current is expected to occur, leading to a finite resistance \cite{Mikitik2001August,Risse1997June}. We applied this configuration to a Nb$_3$Sn single crystal and report on the novel option to unveil the peak effect in the resistivity data by the continuous application of a shaking field $h_{ac}$ parallel to the main magnetic field $H$.

\section{EXPERIMENT}

The Nb$_3$Sn single crystal used for this study (mass $m \approx 11.9\, $mg, thickness $d \approx 0.4\, $mm, and cross section $A \approx 0.44\, $mm$^2$) was characterized by Toyota \emph{et al.}~\cite{Toyota1988September} and was further investigated by Lortz \emph{et al.}~for calorimetric experiments \cite{Lortz2006September,Lortz2007April,Lortz2007March}. The transition to superconductivity in zero magnetic field as determined by a resistivity measurement occurs at $T_{c} \approx 18.0\, $K.\\
\indent Resistivity measurements were performed in a commercial physical property measurement system (PPMS), Quantum Design in a four-point configuration. These measurements were performed in the so-called ac mode, where a dc current abruptly switches its direction $15$ times/s (square-wave excitation). The external main magnetic field was applied perpendicular to the flat side of the crystal $(H \parallel c)$ and also perpendicular to the transport current (see Fig.~\ref{fig.PRB0}(a)).\\
\indent We mounted an ac shaking coil with $1306$ windings on a commercial PPMS resistivity puck with its axis aligned with the main magnetic field so that $h_{ac} \parallel H$. In order to monitor the amplitude of the shaking field $h_{ac}$ we installed an additional pick-up coil inside this ac shaking coil. To achieve an accurate measurement of the sample temperature, we attached a calibrated thermometer (Cernox, LakeShore Inc.) at the top of a brass block supporting the sample. The ac shaking coil was driven by an external ac power source (Agilent model $6811$B).\\
\indent Additional dc magnetization measurements were performed in the same PPMS using the commercial standard option.

\section{RESULTS}

\begin{figure}
\includegraphics[width=75mm,totalheight=200mm,keepaspectratio]{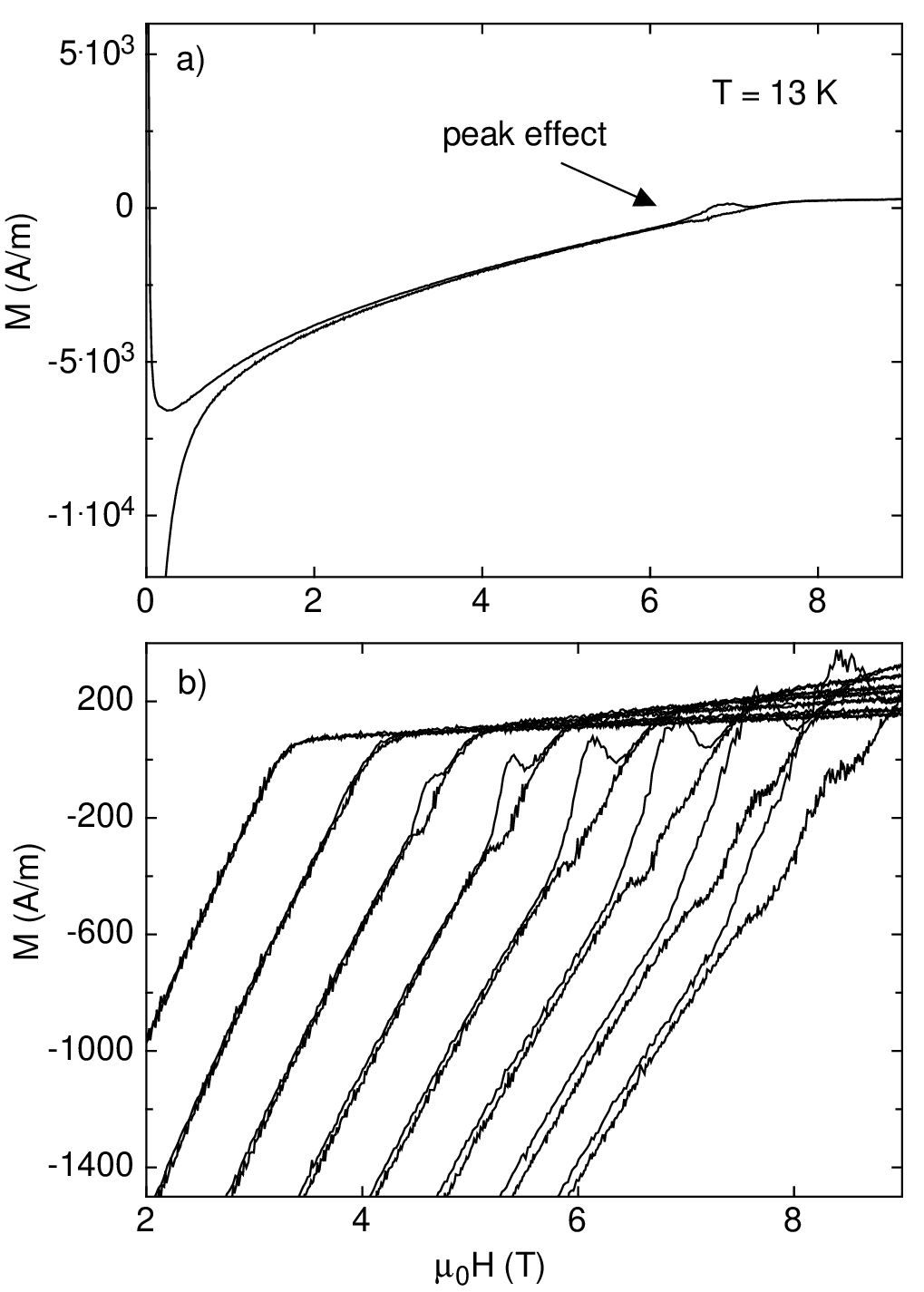}
\caption{\textbf{(a)} Dc magnetization-hysteresis curve for $T = 13\, $K. \textbf{(b)} Parts of the dc magnetization hysteresis curves for different temperatures (from left to right) : $T = 15.5\, $K, $15\, $K, $14.5\, $K, $14\, $K, $13.5\, $K, $13\, $K, $12.5\, $K, $12\, $K. \label{fig.PRB3}}
\end{figure}

\begin{figure}
\includegraphics[width=75mm,totalheight=200mm,keepaspectratio]{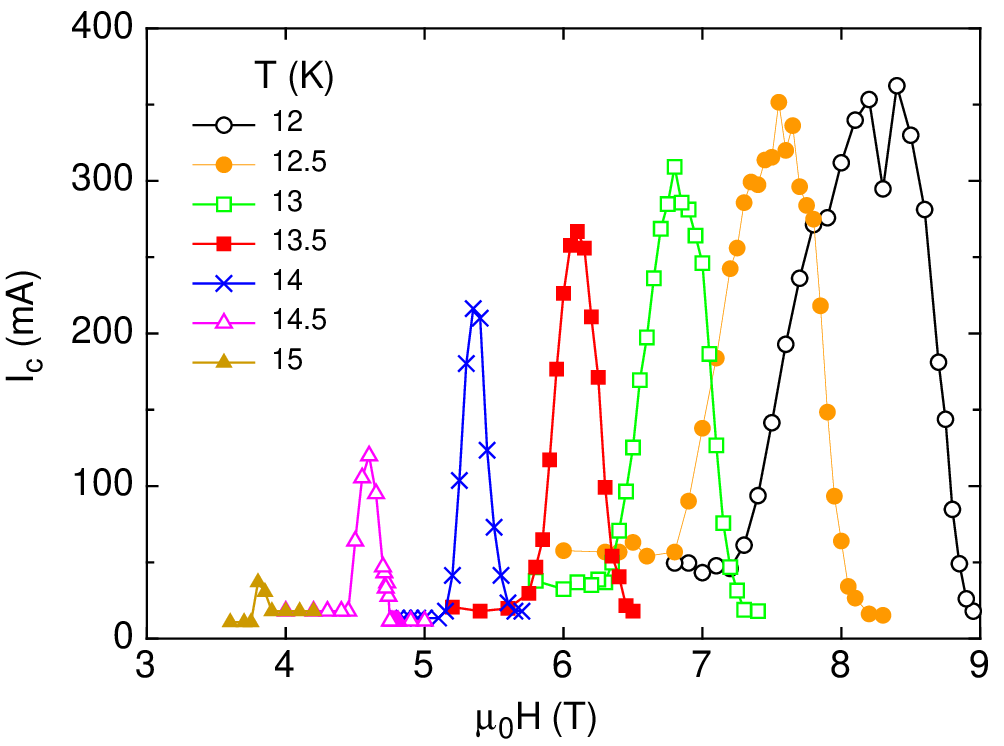}
\caption{(Color online) $I_c(B)$ data as obtained from the magnetization measurements shown in Figs.~\ref{fig.PRB3}. \label{fig.PRB18}}
\end{figure}

\begin{figure}
\includegraphics[width=75mm,totalheight=200mm,keepaspectratio]{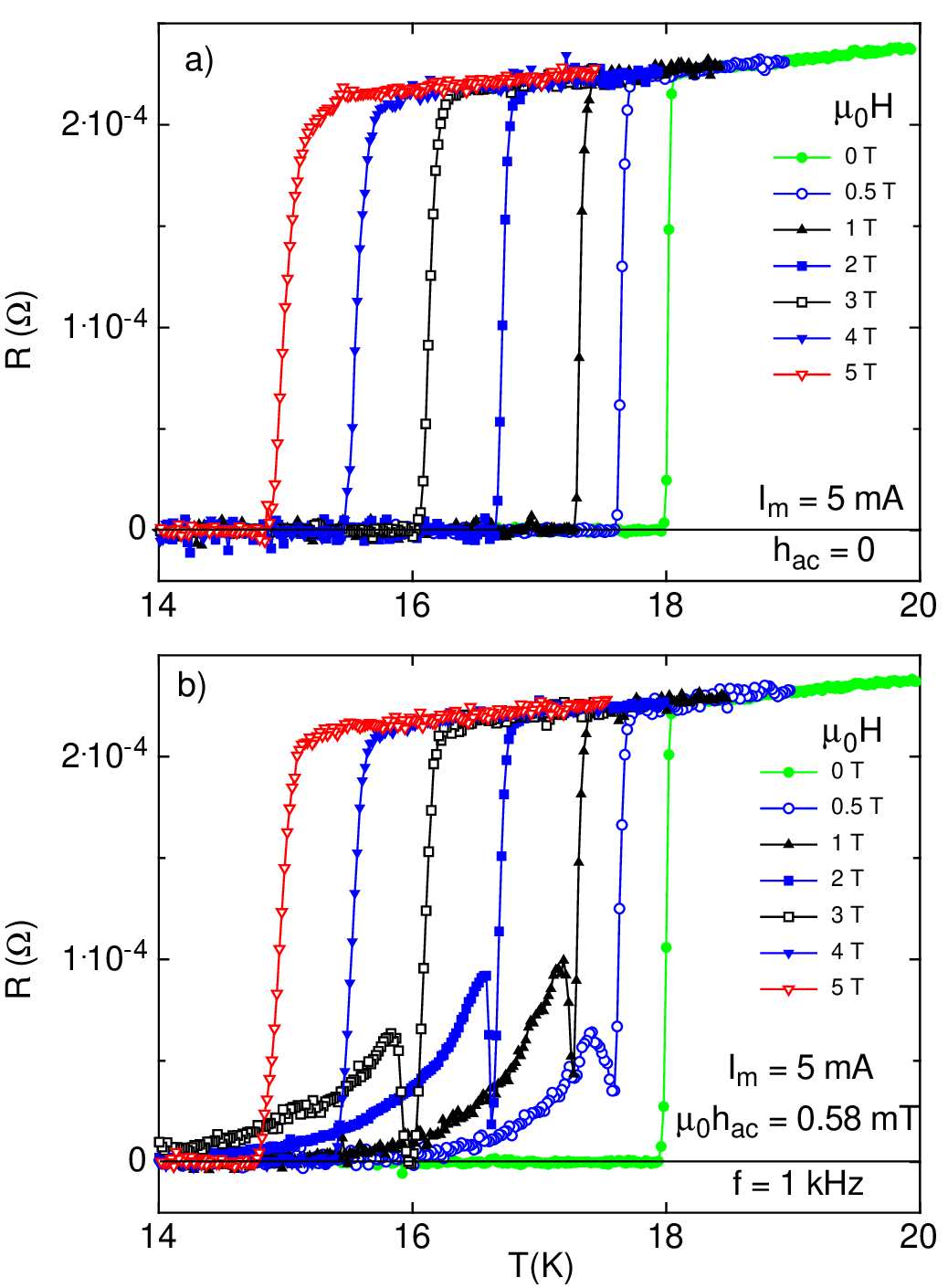}
\caption{(Color online) Resistance $R(T)$ of a Nb$_3$Sn single crystal for different magnetic fields $H \parallel c$, measured with a transport current $I_m = 5\, $mA. The temperature was swept downwards with $dT/dt = - 50\, $mK/min.  \textbf{(a)} Without an oscillating magnetic field, i.e., $h_{ac} = 0$.  \textbf{(b)} With an oscillating magnetic field ($f = 1\, $kHz, $\mu_0 h_{ac} = 0.58\, $mT) superimposed. \label{fig.PRB4}}
\end{figure}

\begin{figure}
\includegraphics[width=75mm,totalheight=200mm,keepaspectratio]{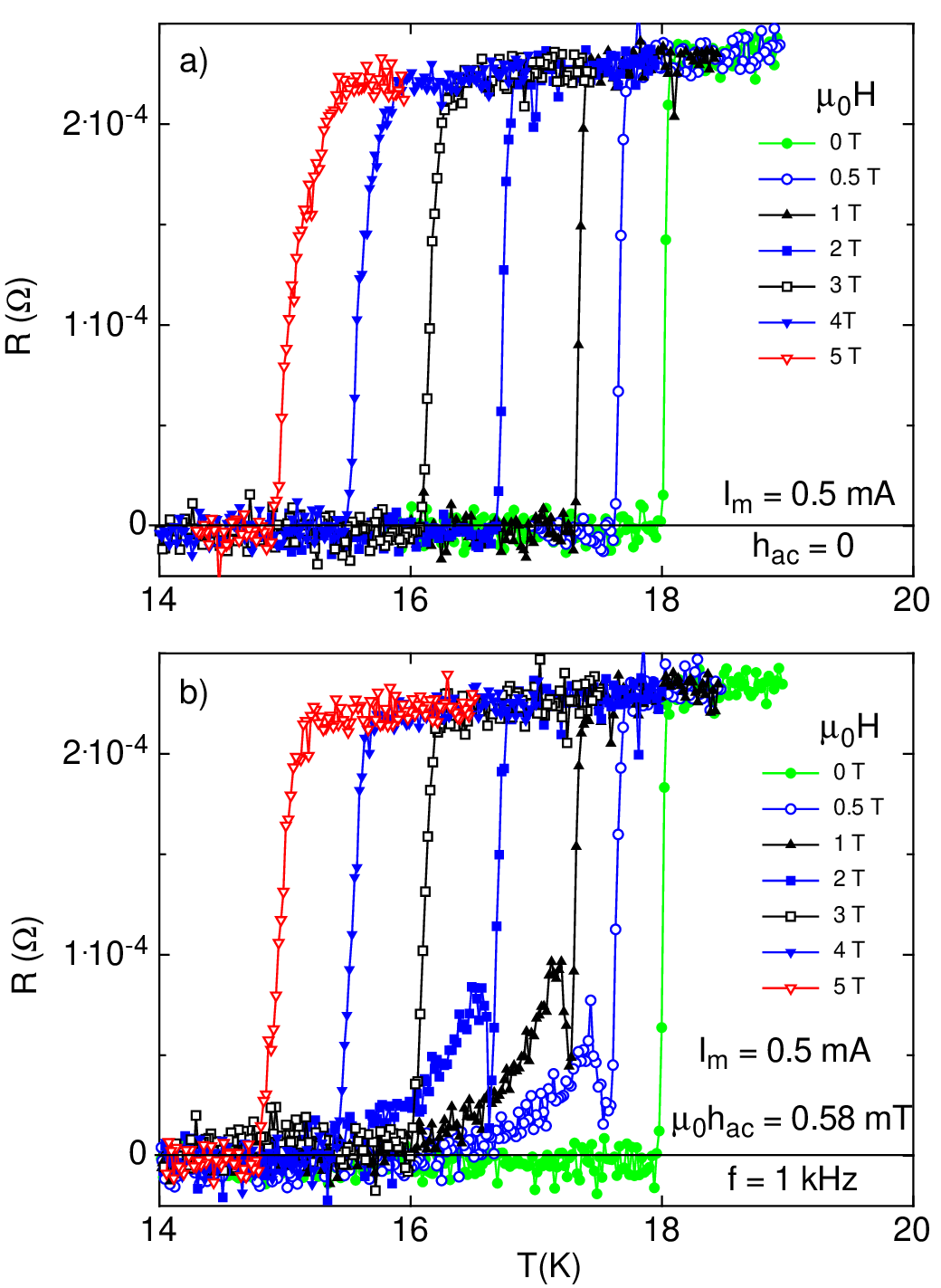}
\caption{(Color online) Resistance $R(T)$ of a Nb$_3$Sn single crystal for different magnetic fields, measured with $I_m = 0.5\, $mA. \textbf{(a)} Without an oscillating magnetic field, i.e., $h_{ac} = 0$.  \textbf{(b)} With an oscillating magnetic field ($f = 1\, $kHz, $\mu_0 h_{ac} = 0.58\, $mT) superimposed. \label{fig.PRB5}}
\end{figure}

\begin{figure}
\includegraphics[width=75mm,totalheight=200mm,keepaspectratio]{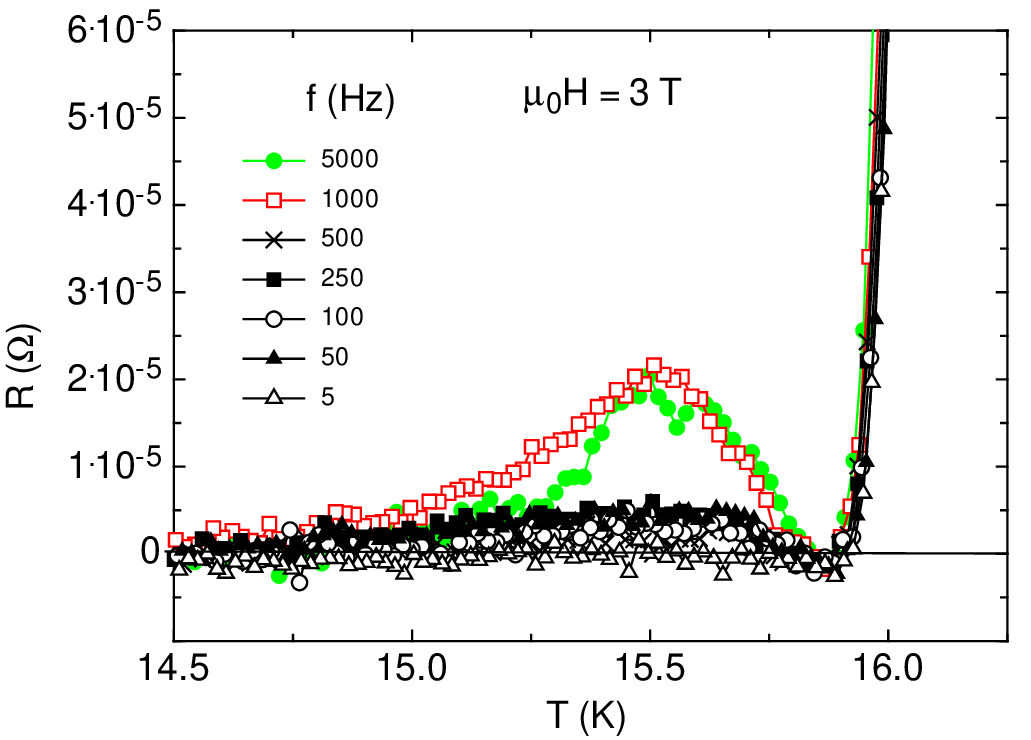}
\caption{(Color online) Temperature sweeps $R(T)$ for $\mu_0 H = 3\, $T, fixed $h_{ac} = 0.29\, $mT and $I_m = 5\, $mA, but different "shaking" frequencies $f$. \label{fig.PRB11}}
\end{figure}

\begin{figure}
\includegraphics[width=75mm,totalheight=200mm,keepaspectratio]{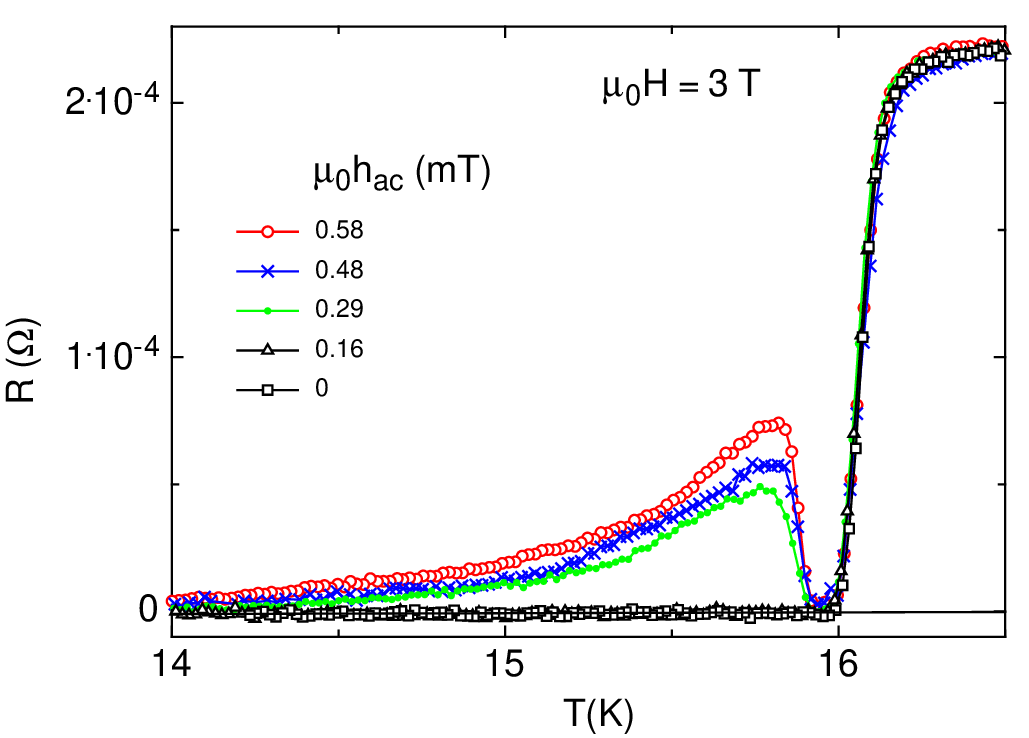}
\caption{(Color online) Temperature sweeps $R(T)$ for $\mu_0 H = 3\, $T, fixed frequency $f = 1\, $kHz, $I_m = 5\, $mA, but different amplitudes of the oscillating magnetic field $h_{ac}$. \label{fig.PRB12}}
\end{figure}

\begin{figure}
\includegraphics[width=75mm,totalheight=200mm,keepaspectratio]{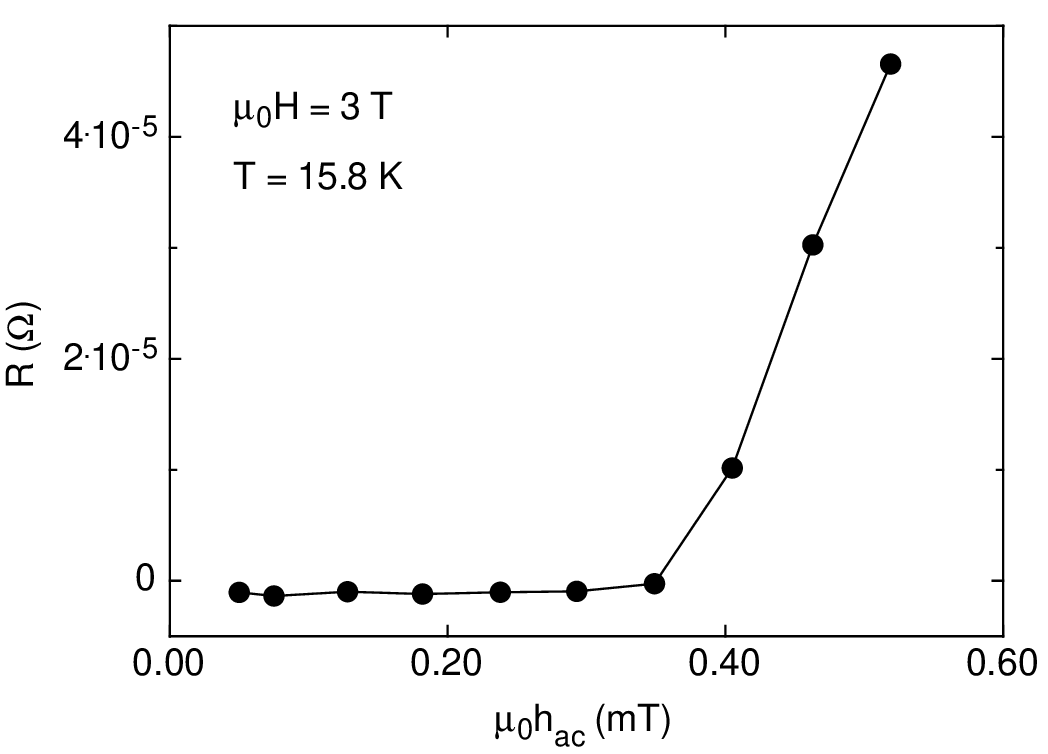}
\caption{Resistance $R$ for different amplitudes of the oscillating magnetic field $h_{ac}$ in $\mu_0 H = 3\, $T, $T = 15.8\, $K, $I_m = 5\, $mA and fixed frequency $f = 1\, $kHz. \label{fig.PRB13}}
\end{figure}

\begin{figure}
\includegraphics[width=75mm,totalheight=200mm,keepaspectratio]{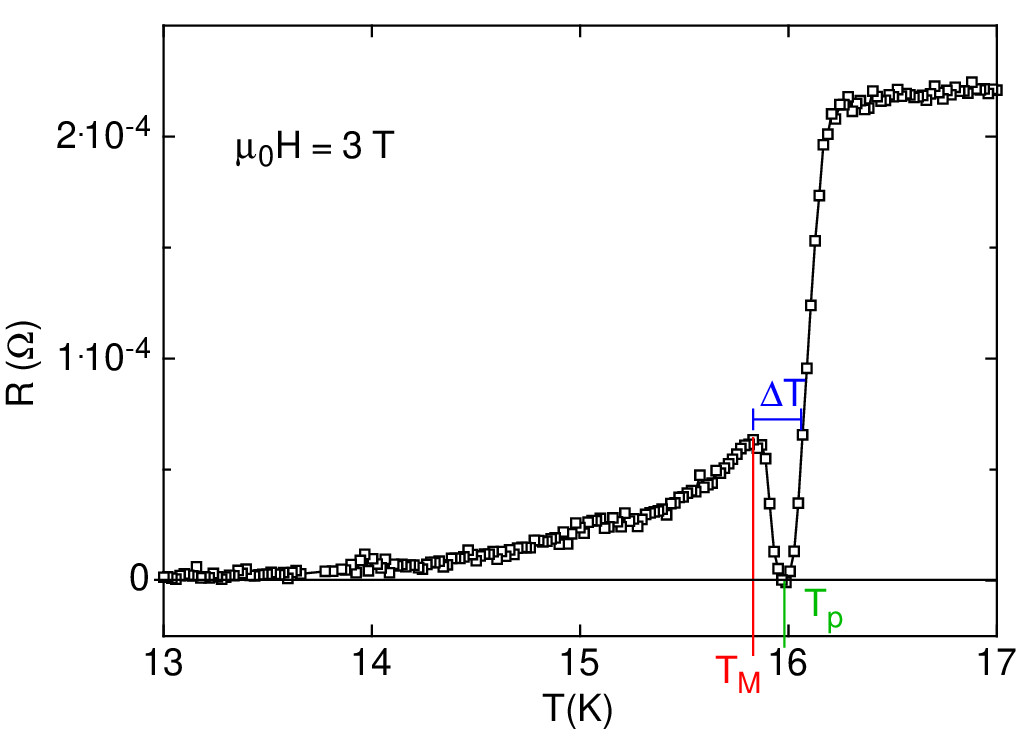}
\caption{(Color online) Temperature dependence of the resistance $R(T)$ for a fixed external magnetic field $\mu_0 H = 3\, $T, with an oscillating magnetic field ($f = 1\, $kHz, $\mu_0 h_{ac} = 0.58\, $mT) superimposed and $I_m = 5\, $mA. \label{fig.PRB6}}
\end{figure}

Earlier investigations on the same Nb$_3$Sn crystal have shown that it exhibits a pronounced magnetic peak effect near $H_{c2}$ \cite{Lortz2007March}. We have performed additional dc magnetization measurements on this crystal and show in Figs.~\ref{fig.PRB3}(a) and \ref{fig.PRB3}(b) corresponding magnetization hysteresis loops for the crystal taken at different temperatures. The bubble-like feature near the upper critical field is caused by the peak effect. In Fig.~\ref{fig.PRB18} we have plotted the critical currents $I_c$ in the peak-effect region as estimated from the magnetization-hysteresis loops shown in Fig.~\ref{fig.PRB3}(b) by using Bean's critical-state model.\\
\indent Figures \ref{fig.PRB4}(a) and \ref{fig.PRB4}(b) show the temperature dependence of the resistance of the same sample for different magnetic fields $H$, with a used transport current $I_m = 5\, $mA. While the data shown in Fig.~\ref{fig.PRB4}(a) were taken without any shaking field $h_{ac}$, the data displayed in Fig.~\ref{fig.PRB4}(b) were measured with $\mu_0 h_{ac} = 0.58\, $mT $\parallel H$. With $h_{ac} = 0$, no feature other then the sharp transition to superconductivity is discernible. This observation indicates that no significant depinning occurs for $T<T_c(H)$ with a transport current $I_m = 5\, $mA. With an additional shaking field $h_{ac}$ of frequency $f = 1\, $kHz parallel to the main magnetic field, ($h_{ac} \parallel H$), however, a clear signature due to the peak effect appears for $\mu_0 H \leq 3\, $T, which is the central result of this paper. The data taken in larger magnetic fields remain unaffected. This signature consists of a region with finite resistance for temperatures $T$ well below $T_c$ which we may attribute to the dissipative motion of magnetic vortices, and a pronounced drop right below $T_c$ that must indicate enhanced flux pinning in the peak-effect region (see discussion below).\\
\indent In Figs.~\ref{fig.PRB5} we show similar data as in Figs.~\ref{fig.PRB4}, but with the transport current reduced by an order of magnitude to $I_m = 0.5\, $mA.\\
\indent We have also systematically studied the dependence of $R(T)$ on the frequency $f$ of the shaking field. In Fig.~\ref{fig.PRB11} we show corresponding temperature sweeps $R(T)$ for frequencies between $5\, $Hz and $5\, $kHz. Very strikingly, only frequencies $f \gtrsim 1\, $kHz induced a significant resistance well below $T_c$.\\
\indent We finally investigated the dependence of $R(T)$ on the amplitude $h_{ac}$ of the shaking field at fixed frequency $f = 1\, $kHz (see Fig.~\ref{fig.PRB12}). A significant resistance below $T_c$ appears abruptly above a certain threshold value, $\mu_0h_{ac} \approx 0.3\, $mT. To demonstrate this effect more clearly we have plotted in Fig.~\ref{fig.PRB13} corresponding data taken at a constant $T = 15.8\, $K, where the sample resistance becomes finite at $\mu_0 h_{ac} \approx 0.4\, $mT. The apparent discrepancy between the observed threshold values for $h_{ac}$ appearing in Figs.~\ref{fig.PRB12} and \ref{fig.PRB13} may be a consequence of a certain irreversible behavior around the peak-effect region, so that experiments with varying temperature and corresponding measurements with varying magnetic field (and at fixed $T$) may not give exactly the same results.\\
\indent In order to discuss these features in the $R(T)$ curves that show the peak effect, we plotted in Fig.~\ref{fig.PRB6} a selected $R(T)$ curve for $\mu_0 H = 3\, $T. We define the temperatures $T_p$ as the temperature at the local minimum, $T_M$ as the temperature at the local maximum of the $R(T)$ curve, respectively, and $\Delta T$ as the distance on the temperature scale between the local maximum of the resistance at $T_M$ and the sharp increase in $R(T)$ at the transition to the normal state near $T_c$ (see Fig.~\ref{fig.PRB6}).

\section{DISCUSSION}

\begin{figure}
\includegraphics[width=75mm,totalheight=200mm,keepaspectratio]{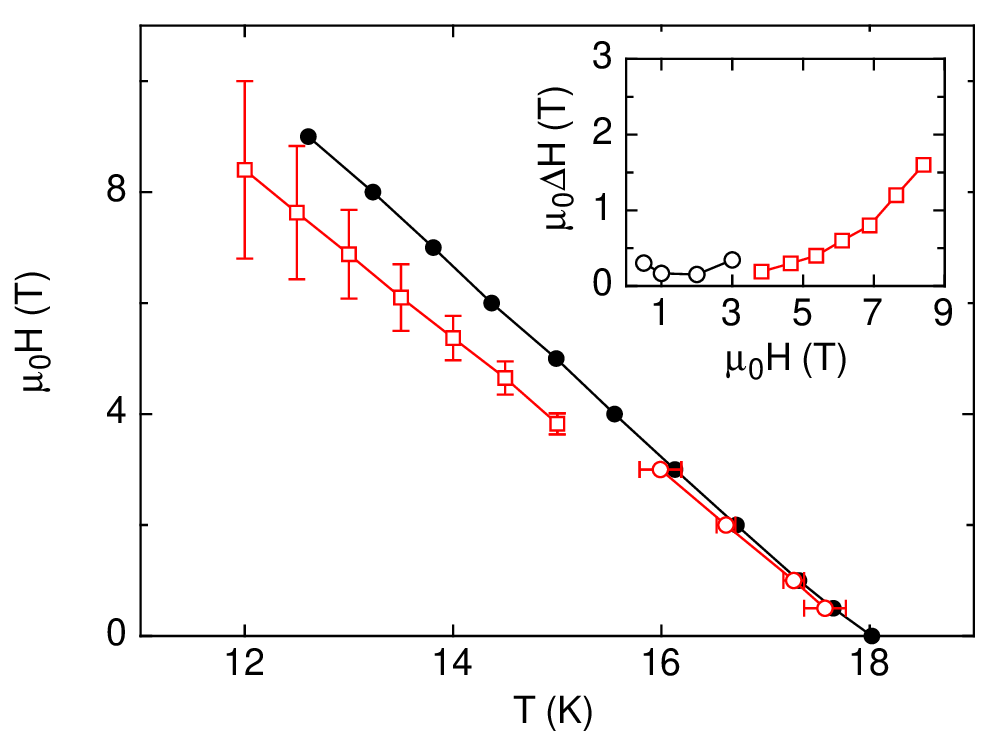}
\caption{(Color online) Part of the magnetic phase diagram of the Nb$_3$Sn single crystal as derived from the data shown in Figs.~\ref{fig.PRB3} and \ref{fig.PRB4}. Filled circles : upper critical field; open circles : $T_p$; open squares : center of the peak-effect region from $M(H)$ data. Inset : magnetic-field dependence of the width of the peak-effect region $\Delta H$. Circles represent data as deduced from the resistivity measurements of Fig.~\ref{fig.PRB4}(b), while squares stand for data extracted from the dc magnetization measurements shown in Figs.~\ref{fig.PRB3}. \label{fig.PRB15}}
\end{figure}

A minimum in the resistance $R(T)$ at $T_p$ close to $T_c$, as observed in electrical transport measurements on, e.g., V$_3$Si \cite{Chaudhary2001,Meier-Hirmer1985January}, Nb$_3$Ge, Mo$_3$Si \cite{Wordenweber1986March}, $2$H-NbSe$_2$ \cite{Higgins1996,Henderson1996September}, Nb \cite{Daniilidis2007,Huebener1970February}, CeRu$_2$, CeCo$_2$, Yb$_3$Rh$_4$Sn$_{13}$ \cite{Sato1997}, and YBa$_2$Cu$_3$O$_7$ \cite{Langan1997}, has usually been interpreted as a manifestation of the peak effect. In all these cited experiments, such a minimum was observed without the aid of a superimposed ac field. Only in Ref.~\cite{Chaudhary2001} the authors used a configuration comparable to ours by cycling of the main magnetic field prior to their measurements on V$_3$Si in order to restore the signature of the peak effect. In our measurements on Nb$_3$Sn, however, such a characteristic feature in $R(T)$ appears by the continuous application of a shaking field $h_{ac}$, which allows us to restore the peak effect in a very convenient and efficient way with a high data-point density. This shaking field $h_{ac}$ exerts a periodic Lorentz force on the magnetic flux lines, in combination with an additional constant force induced by the used transport current $I_m$. As soon as these combined forces exceed the pinning force, the flux lines will be depinned and the resulting dissipative motion causes a finite resistance for $T<T_c$. According to an explanation of the peak effect by Pippard \cite{Pippard1969}, the FLL softens in the peak-effect region, i.e., the shear modulus is vastly reduced, the FLL becomes less rigid, and the flux lines can bend and adjust better to the pinning sites. Since the resulting pinning is stronger in the peak-effect region, the motion of the flux lines is hindered and therefore the resistance decreases. This increased pinning strength leads to an increase in the critical current, which can also be observed as a drop in the real part of the ac susceptibility (see, for example, Refs.~\cite{Marchevsky2001February} and \cite{Adesso2006March}).\\
\indent To prove that the observed features in $R(T)$ at $T_p$ coincide with the magnetically determined peak effect, we show in Fig.~\ref{fig.PRB15} the results from Figs.~\ref{fig.PRB3}(b) and \ref{fig.PRB4} in one magnetic phase diagram. The lines correspond to the position of the center of the peak-effect region and seem to match for both data sets, which strongly supports the interpretation of the observed drop in resistance at $T_p$ as a signature of the peak effect. The vertical and horizontal bars drawn in this figure are derived from the widths of the respective features that we ascribe to the peak effect. The inset of Fig.~\ref{fig.PRB15} shows the magnetic-field dependence of the width $\Delta H$ on the magnetic-field scale for both data sets, where we have converted $\Delta T$ according to $\Delta H = \Delta T dH_{c2}/dT$. The $\Delta H$ as derived from the resistivity data also essentially coincide with the corresponding widths extracted from the dc magnetization data.\\
\indent Adesso \emph{et al.}~\cite{Adesso2006March} investigated the same Nb$_3$Sn crystal by studying the higher harmonics of the ac magnetic susceptibility $\chi_{ac}$. The peak effect was detected in magnetic fields between $3$ and $13\, $T, and the resulting magnetic phase diagram is in line with ours (Fig.~\ref{fig.PRB15}).\\
\indent In the following we discuss the dependence of the observed features in $R(T)$ on the transport current $I_m$, on the oscillation frequency $f$, and on the amplitude $h_{ac}$. Reducing the transport current to $I_m = 0.5\, $mA leads naturally to more scattered data in Figs.~\ref{fig.PRB5} as compared to the measurements for $I_m = 5\, $mA shown in Figs.~\ref{fig.PRB4}. Noteworthy, flux-line motion sets in at $\mu_0 H = 2\, $T and below for $I_m = 0.5\, $mA (Fig.~\ref{fig.PRB5}(b)), while it already occurs in $\mu_0 H = 3\, $T for the higher transport current $I_m = 5\, $mA (Fig.~\ref{fig.PRB4}(b)). This trend can be explained if one assumes, in line with the Anderson-Kim flux creep theory \cite{Anderson1962}, that a large transport current leads to a higher mobility of flux-line bundles due to the stronger Lorentz force as compared to a situation with a small current $I_m$.\\
\indent The existence of a threshold frequency above which a finite resistance below $T_c$ abruptly sets in seemingly deviates from the findings by Risse \emph{et al.}~\cite{Risse1997June} on amorphous Mo$_3$Si films, where a linear dependence of the resistance below $T_c$ on the frequency of the shaking field was reported, without any threshold frequency. We note, however, that while the maximum amplitude of the shaking field in our measurements, $\mu_0 h_{ac} \approx 0.58\, $mT, was comparable to that used by Risse \emph{et al.}~($0.47\, $mT) \cite{Risse1997June}, the frequencies used in Ref.~\cite{Risse1997June} were usually between $100\, $kHz and $300\, $kHz, much higher than those used in our experiment ($\leq 5\, $kHz, $f = 1\, $kHz for most experiments). Therefore, a possible threshold frequency may have gone unnoticed in Ref.~\cite{Risse1997June}, and/or the different material parameters and degree of disorder may play a certain role here.\\
\indent We now want to compare our results with the estimates made in Ref.~\cite{Mikitik2001August} for the $h_{ac} \parallel H$ configuration. The vortex system can be equilibrated if the amplitude $h_{ac}$ of a shaking field fulfills $h_{ac} \geq h^* =(1-I/I_c)h_p$, with $h_p=j_cb/2$. For too small values of $h_{ac}$ the shaking field cannot fully penetrate the sample, and parts of the sample remain unaffected by the ac magnetic field. Such an argument may explain the observation of a threshold value $h^*$ for $h_{ac}$ above which the peak effect can be observed in our $R(T)$ data. To make a quantitative comparison with the above prediction we assumed the validity of Bean's critical-state model, where $j_c$ is assumed to be spatially constant. We estimated $I_c \approx 27\, $mA and $j_c \approx 6.1 \times 10^4\, $A/m$^2$ from a corresponding measurement of the current-voltage characteristics in $\mu_0 H = 3\, $T and at $T = 15\, $K and using a voltage criterion $V \lesssim 0.1\, \mu$V. With the sample width $b \approx 1.1\, $mm and the measuring current $I= 5\, $mA, we obtain the condition $\mu_0 h_{ac} \geq \mu_0 h^* = 0.035\, $mT in $\mu_0 H = 3\, $T and at $T = 15\, $K. This value is an order of magnitude smaller than our measured threshold value, $\mu_0 h^* \approx 0.3 - 0.4\, $mT for an $h_{ac} \parallel H$ configuration as used in our experiment (see Figs.~\ref{fig.PRB12} and \ref{fig.PRB13}).\\
\indent We have used the relations from Ref.~\cite{Mikitik2001August} to estimate the time-averaged electric field $E_{av}$ and the resulting resistance $R = l\, E_{av}/I$ along the sample length $l \approx 3.3\, $mm. From $E_{av} = \frac{I}{I_c} 2bf \mu_0(h_{ac}-h^*)$ for $h_{ac} > h^*$, with $I = 5\, $mA and $f = 1\, $kHz, we obtain $R \approx 130\, \mu\Omega$ for $\mu_0 H = 3\, $T and $T = 15\, $K. The measured value $R \approx 30\, \mu\Omega$ is lower than this estimated value. In addition, considering the fact that corresponding measurements with $f = 5\, $kHz give almost the same $R$ values, we cannot confirm the predicted linear dependence of $R$ on $f$ in Ref.~\cite{Mikitik2001August}.\\
\indent We note that the effects of ac transport currents on the flux-line lattice can be similar to those of a superimposed shaking field produced by an ac coil as it has been shown by Gittleman and Rosenblum \cite{Gittleman1966April} who used radio frequencies for their transport current, but without investigating the peak effect any further. Higgins and Bhattacharya \cite{Higgins1996} measured the peak effect using an ac source for the transport current ($I_m \sim 5\, $mA, unknown frequency) with a standard lock-in technique, but without the aid of a superimposed shaking field.\\
\indent In our experiment, the square-wave excitation of the measuring current corresponds to a leading measuring frequency $f_1 = 7.5\, $Hz, with rapidly decreasing amplitude of the higher harmonics. Using the total current $I = 5\, $mA and applying Ampere's law (with the sample approximated by a cylinder of equal cross section $A \approx 0.44\, $mm$^2$), we obtain a self field at the sample surface of the order of $3\, \mu$T. Both the used measuring frequency and this self field are well below the here reported threshold values ($500\, $Hz and $160\, \mu$T, respectively), above which the shaking has shown to become effective, and we therefore conclude that the influence of the measuring current is negligible when compared to that of the external main shaking field.\\
\indent We finally want to point out that the technique of measuring $R(T)$ with an external continuously oscillating magnetic field may be more sensitive to explore the peak effect than inferring it from corresponding magnetization $M(H)$ measurements alone, at least in some parts of the magnetic phase diagram of a sample. While the peak-effect feature is not discernible at temperatures higher than $15\, $K in the $M(H)$ data shown in Figs.~\ref{fig.PRB3}, the resistivity data (taken in combination with an ac magnetic field) reveal the peak-effect region also in this high-temperature region of the magnetic phase diagram. The absence of a corresponding signature in our resistivity data for magnetic fields larger than $3\, $T, however, can be explained with a strongly increasing $j_c$ with decreasing temperature, so that a condition similar to $h_{ac} \geq j_cb/2$ can no longer be fulfilled due to technical limitations for the amplitude $h_{ac}$. We expect, however, that larger shaking amplitudes $h_{ac}$ do, in principle, allow to study the peak effect also in higher magnetic fields and at lower temperatures.

\section{CONCLUSION}

We have reported on the observation of the manifestation of the peak effect in resistivity data that is induced by the continuous application of an external oscillating magnetic field $h_{ac} \parallel H$. No corresponding signature appears in the $R(T)$ data of the Nb$_3$Sn crystal that were taken without such a shaking magnetic field.\\
\indent We have studied the dependence of this effect on the transport current, the shaking frequency $f$, and the amplitude $h_{ac}$ of the oscillating magnetic field. The amplitude $h_{ac}$ that has been necessary to unveil the peak effect is an order of magnitude larger than what has been predicted to bring the vortex system to an equilibrium. These resistivity measurements, combined with a small oscillating magnetic field, allowed us to detect the signature of the peak effect also in the high-temperature part of the magnetic phase diagram of Nb$_3$Sn where dc magnetization data were not accurate enough. Therefore we believe that such resistivity measurements, in combination with an oscillating magnetic field, may turn out to be a versatile tool not only for the investigation of the peak effect but also for studying other effects where the pinning of magnetic flux lines plays a certain role.\\

\section{ACKNOWLEDGMENT}

This work was supported by the Schweizerische Nationalfonds zur
F\"{o}rderung der Wissenschaftlichen Forschung, Grants No.~$20$-$111653$ and No.~$20$-$119793$.

\bibliographystyle{apsrev}

\end{document}